# Dynamic Shimming in the Cervical Spinal Cord for Multi-Echo Gradient-Echo Imaging at 3 T

E. Alonso-Ortiz[1*], D. Papp[1], A. D'Astous[1], J. Cohen-Adad[1,2,3,4]

[1]NeuroPoly Lab, Institute of Biomedical Engineering, Polytechnique Montréal, QC, Canada
[2] Functional Neuroimaging Unit, CRIUGM, Université de Montréal, Montréal, QC, Canada
[3] Mila – Quebec AI Institute, Montréal, QC, Canada
[4] Centre de recherche du CHU Sainte-Justine, Université de Montréal, Montréal, QC, Canada

*Corresponding Author

**Abstract**

Obtaining high quality images of the spinal cord with MRI is difficult, partly due to the fact that the spinal cord is surrounded by a number of structures that have differing magnetic susceptibility. This causes inhomogeneities in the magnetic field, which in turn lead to image artifacts. In order to address this issue, linear compensation gradients can be employed. The latter can be generated using an MRI scanner's first order gradient coils and adjusted on a per-slice basis, in order to correct for through-plane ("z") magnetic field gradients. This approach is referred to as z-shimming. The aim of this study is two-fold. The first aim was to replicate aspects of a previous study wherein z-shimming was found to improve image quality in T2*-weighted echo-planar imaging. Our second aim was to improve upon the z-shimming approach by including in-plane compensation gradients and adjusting the compensation gradients during the image acquisition process so that they take into account respiration-induced magnetic field variations. We refer to this novel approach as realtime dynamic shimming. Measurements performed in a group of 12 healthy volunteers at 3 T show improved signal homogeneity along the spinal cord when using z-shimming. Signal homogeneity may be further improved by including realtime compensation for respiration-induced field gradients and by also doing this for gradients along the in-plane axes.

## Introduction

An important challenge for spinal cord (SC) and brain imaging is proximity of the tissues of interest to air in the lungs, to cartilage, and to bone. The different magnetic susceptibility ($\chi$) between soft tissues in the brain and SC and these surrounding structures leads to macroscopic (greater than the voxel dimensions) static field inhomogeneities ($\Delta B_0$), which can result in signal loss in $T_2^*$-weighted images (Finsterbusch, Eippert, and Büchel 2012) and geometric distortions in echo-planar imaging (EPI) (Jezzard and Balaban 1995). Further compounding this issue is the fact that during respiration, the volume of air in the lungs changes, leading to time-varying $\Delta B_0$.

Static $\Delta B_0$ can be minimized by active shimming; generating compensatory magnetic fields with approximately the same spatial distribution and magnitude, but opposite sign, as that of $\Delta B_0$. Shimming is commonly achieved by measuring $\Delta B_0$, and subsequently decomposing the $\Delta B_0$ distribution into a set of spherical harmonic basis functions. Three first order spherical harmonic terms describe linear deviations from a uniform field, corrected using linear gradient coils. The five second order spherical harmonic terms describe quadratic deviations from a uniform field, which are generated using a combination of the linear gradient coils and additional second order coils in clinical/typical scanners (Stockmann and Wald 2018). However, in areas of strong, localized $\Delta B_0$, such as the SC region, the prefrontal cortex or the medial temporal lobe, severe residual inhomogeneities remain after conventional shimming (Finsterbusch, Eippert, and Büchel 2012). To address this issue, a number of solutions have been proposed, such as higher-order spherical harmonic insert coils (Punchard et al. 2013), external multi-coil arrays that produce non-orthogonal field patterns (Topfer et al. 2016; Juchem et al. 2011), or "dynamic shimming", whereby spherical harmonic shims can be optimized on a slice-by-slice basis (Juchem et al. 2011; Blamire, Rothman, and Nixon 1996).

MRI of the SC often necessitates high in-plane resolution, due to the small dimensions of the SC structures, and thick slices, to maintain an adequate signal-to-noise (SNR) ratio. Consequently, sensitivity to $\Delta B_0$ has often been considered to be highest along the slice-select direction. In order to reduce the effects of high $\Delta B_0$ along the slice-select direction, a technique called "z-shimming" has been used for brain MRI. Z-shimming seeks to reduce through-slice signal dephasing in $T_2^*$-weighted acquisitions by applying slice-specific linear compensation gradients along the slice-select axis (Constable and Spencer 1999). Z-shimming has been successfully employed by Finsterbusch et al. (Finsterbusch, Eippert, and Büchel 2012) in the cervical SC using single slice-specific compensation gradients, thanks to the low in-plane variability of $\Delta B_0$ in this small region-of-interest (ROI). In (Finsterbusch, Eippert, and Büchel 2012), the z-shimming approach was applied to $T_2^*$-weighted EPI axial acquisitions of the cervical spinal cord. Results obtained in 24 healthy volunteers at 3 T indicated that z-shimming increased signal intensities in the spinal cord (overall by about 20%) and reduced signal intensity variations between different slices along the spinal cord (by about 80%).

Z-shimming does not however compensate for time-varying effects. In order to mitigate the effects of time-varying $\Delta B_0$, a number of solutions have been proposed, such as, navigator echoes (Barry and Menon 2005; Pfeuffer et al. 2002), retrospective correction methods (Glover, Li, and Ress 2000; Vannesjo et al. 2019) or real-time (i.e. "realtime") control shimming. The latter has been demonstrated in the brain at 7 T by monitoring respiration with respiratory bellows and adjusting the scanner's 1st and 2nd order spherical harmonic shims in real-time (van Gelderen et al. 2007). In (van Gelderen et al. 2007), hardware modifications were made so that the shims could be run independently of the scanner. More recently, realtime shimming was achieved in the SC using a custom-built shim coil (Topfer et al. 2018). While the prospective nature of realtime shimming is attractive, its reliance on specialized hardware limits its widespread adoption.

In this work, we sought to reproduce previous attempts to improve upon $T_2^*$-based cervical spinal cord images using z-shimming (Finsterbusch, Eippert, and Büchel 2012) by implementing z-shimming within a multi-echo gradient-echo (MGRE) sequence. To expand upon this work we included in-plane (to correct for magnetic field gradients along the frequency and phase encode directions) compensation gradients within our MGRE sequence and investigated a solution for the time-varying component of $\Delta B_0$ in which compensation gradients are updated on a per-slice basis and in real-time to reflect respiration-induced changes in ΔB$_0$. The approach, which we call realtime dynamic shimming, does not require hardware modifications, or specialized coils.

**Theory and Background**

We assume that during normal breathing, the magnetic field at every point in space varies sinusoidally in time and oscillates about an equilibrium value (ΔB$_{static}$) with an angular frequency ω$_{resp}$ (Topfer et al. 2018; Vannesjo et al. 2019). We do not expect this assumption to hold during breath-hold experiments, or in other irregular breathing scenarios. The magnetic field offset (ΔB$_0$ (t)) within a given voxel can therefore be expressed as:

$$\Delta B_0(t) = \Delta B_{static} + RIRO_{max} sin(\omega_{resp} t) \qquad [1]$$

where *RIRO$_{max}$* is the maximum respiration-induced resonance offset. Hereafter, values of ΔB$_0$ are considered to be in units of Hz, following common convention.

If one assumes that the magnetic field varies linearly across a voxel, the magnetic field offset within a given voxel can be approximated as:

$$\Delta B_0(t) = G_{x,static}\Delta x + G_{y,static}\Delta y + G_{z,static}\Delta z$$

$$+ (RIGO_{x,max}\Delta x + RIGO_{y,max}\Delta y + RIGO_{z,max}\Delta z) \cdot sin(\omega_{resp} t) \qquad [2]$$

Where *RIGO$_{i,max}$*, the maximum respiration-induced gradient offset in the i = x,y,z direction is

*RIGO$_i$* = δ*RIRO$_i$*/δi and Δx, Δy, and Δz are the voxel dimensions. It is well known that ΔB$_0$(t) leads to voxel displacements, distortions and signal loss (Yablonskiy et al. 2013), (Reeder et al. 1997).

If *G$_i$* can be assumed to be relatively uniform across an ROI of an imaged slice, its effects can be minimized by including compensation gradients along the "*i*" direction in the pulse sequence used for image acquisition.

In (Finsterbusch, Eippert, and Büchel 2012), the authors aimed to compensate for static magnetic field inhomogeneities along the slice-select direction ($G_z$) in cervical spinal cord imaging by including slice-specific compensation gradients along the slice-select direction ($G_{z,corr}$) in an EPI sequence. The moment (amplitude·duration) of $G_{z,corr}$ should compensate for the dephasing effect of $G_z$ that accumulates during the echo time.

The use of an EPI sequence was motivated by its application to fMRI of the cervical spinal cord. In order to determine the optimal compensation moments, EPI reference scans were acquired during free-breathing with a range of different optimal compensation moments for each imaged slice. In each slice, an ROI was manually drawn in the SC and the maximum EPI signal intensity vs. compensation moment was identified. The latter values were used as compensation moments for each slice of a subsequent EPI scan (a "z-shimmed" EPI scan).

Here we sought to reproduce z-shimming in the cervical SC using an MGRE FLASH-based sequence and to explore additional shimming solutions. In Figure 1 we show the pulse sequence diagram for an MGRE sequence that was modified to include compensation gradients along the through-slice and in-plane directions ($G_{x,corr}$, $G_{y,corr}$, and $G_{z,corr}$). This sequence was also modified such that, at the time of each RF excitation pulse, the pressure measured from respiratory bellows, which were strapped onto the imaged participant, would be measured. With pressure serving as a correlate for the respiratory state (van Gelderen et al. 2007), Eq. 2 can be used to determine the necessary compensation gradients required to correct for static and/or static plus respiratory-induced field inhomogeneities along one or various axes (see Methods for further details).

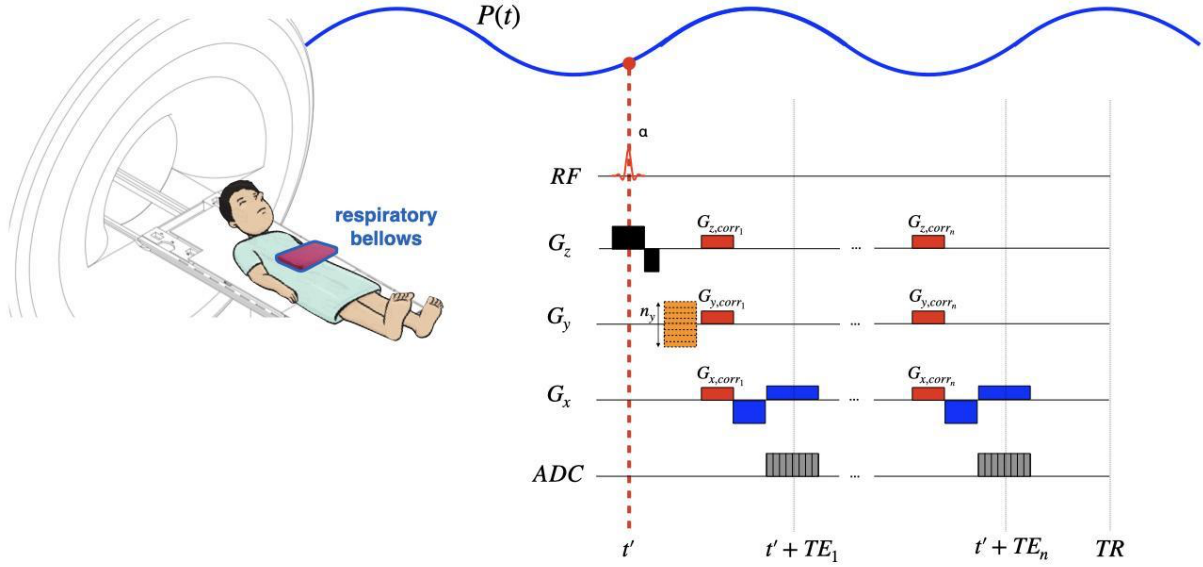

**Figure 1: Modified MGRE sequence.** For each TR and echo time, correction gradients ($G_{corr}$) are played out prior to sampling. The correction gradients are the sum of the static component ($G_{static}$) and the respiration-induced gradient offset (*RIGO*) component, modulated by pressure (*P(t)*) and adjusted for each echo time.

The motivation for selecting an MGRE sequence is due to its interest in quantitative MRI (qMRI). Using an MGRE sequence one can measure $T_2^*$ decay in the spinal cord, opening up a plethora of potential qMRI methods, such as multicomponent $T_2^*$ mapping, $T_2^*$-based myelin water imaging (Alonso-Ortiz, Levesque, and Pike 2018), and quantitative susceptibility mapping (QSM) (Ruetten, Gillard, and Graves 2019), among others. However, $T_2^*$ decay times will be biased by magnetic field inhomogeneities, making these methods highly challenging to apply in regions of strong magnetic field inhomogeneities, such as the spinal cord.

**Methods**

*Phantom*

In order to assess the feasibility of realtime dynamic shimming, we first conducted an experiment using an adapted mechanical phantom (Morozov et al. 2018) to simulate the time-varying $\Delta B_0$ induced by breathing. An air compressor (located in the equipment room) was used to drive a motor (centrifugal fan) which drove two sticks back-and-forth with a period of approximately 2.5 s, close to that of human respiration. A small ferromagnetic object (a staple) was attached to the end of the first stick, which was placed at the edge of a test tube (1.5 cm in diameter) filled with mineral oil. The size of the test tube was carefully selected to ensure that $G_z$ would be as uniform as possible along its cross section. We therefore focused on validating realtime z-shimming with this setup.

The tip of the second stick was placed such that it would push against respiratory bellows that had been taped onto a large, heavy phantom, in an oscillatory fashion (Figure 2). The respiratory bellows were connected to Siemens' physiological monitoring unit (PMU) system, which consists of a transmitter unit that sends the pressure signal, sampled at 50 Hz, to the scanner. The test tube was scanned using a 4-channel small flexible coil on a 3 T Prisma MRI scanner (Siemens Healthineers, Erlangen, Germany).

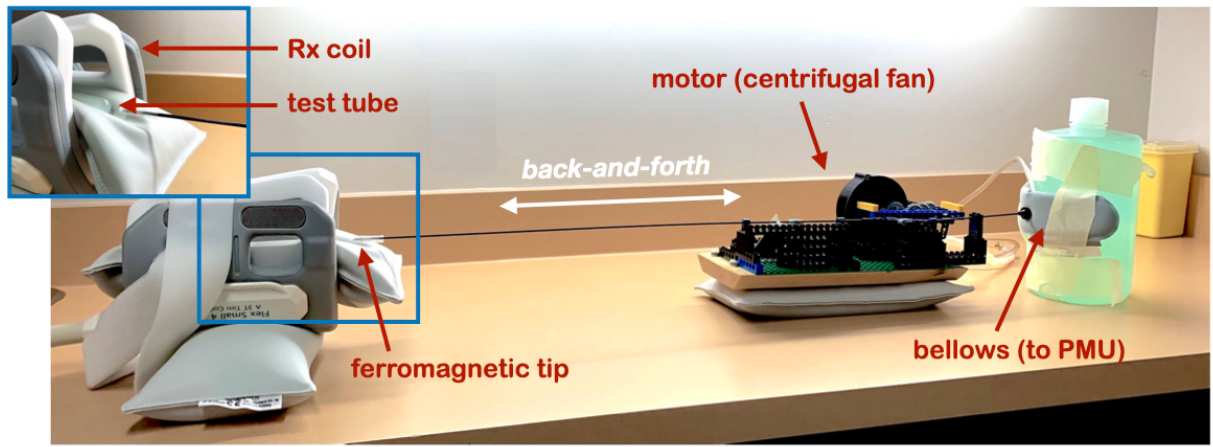

Figure 2: Pneumatic phantom setup. An air compressor blows on a fan, the rotation motion is converted to a sine-like translational motion via a Lego-based system. A long stick presses on respiratory bellows on one end of the stick while the other end has a small ferromagnetic tip attached to it and is located close to the test tube to be imaged.

Data Acquisition: Training Session

The scanner's shim volume was manually set to cover the test tube for volume-specific shimming using the scanner's first and second order shim coils. After shimming, a 1.5 mT/m gradient was generated along the z-axis using the scanner's first order linear shim coil. Next, 60 sequential, sagittal $\Delta B_0$ maps were acquired along the midline of the test tube, using a dual-echo field mapping sequence (TR = 21 ms, TE1 = 2.46 ms, TE2 = 4.92 ms, BW = 900 Hz/pixel, FA = $20^0$, matrix = 96 x 40, ⅞ phase partial Fourier factor, image resolution = 1.25 x 1.25 mm$^2$, 3 contiguous slices of 2.4 mm thickness). At the start of the field map series the PMU signals read into the scanner were automatically written to a log file, along with the log file start and stop times. The table position, the scanner's shim adjustment volume, and the shim values used for $\Delta B_0$ mapping were kept constant throughout all the scans that were subsequently acquired. An axial MGRE scan was acquired perpendicular to the central axis of the test tube (TR = 189 ms, 4 echoes with 3 ms inter-echo spacing, $TE_1$ = 3 ms, BW = 1400 Hz/px, 12 slices contiguous slices of 3 mm thickness, matrix = 128 x 56, image resolution = 1.4 x 1.4 mm$^2$, FA = $25^0$, A/P phase encoding). All the data were transferred to a laptop via an ethernet socket. This scan will be referred to as the "no shim" scan from this point on, to indicate that no dynamic shimming was applied (however, the first and second order volume-specific shim used for the $\Delta B_0$ maps was maintained).

Data Analysis: *Shimming Toolbox*

Once the data were transferred to an external computer, the *Shimming Toolbox* (D'Astous et al. 2021) was used for analysis. The dual-echo field mapping sequence produced phase-difference images which were unwrapped using PRELUDE (Jenkinson 2003). Phase jumps between time points in the time-series of phase-difference images were accounted for by calculating a common validity mask over all timepoints and calculating the mean phase over that mask for each timepoint. If the mean phase between timepoints was greater than pi, we corrected for the n*2pi offset. Next, all phase difference images were converted into field maps ($\Delta B_0$) and smoothed using a Gaussian filter with a standard deviation of 1.5.

The time stamps in the PMU log file were used to associate the pressure readings with the ΔB₀ acquisition times, which are recorded in the header of the phase-difference image files. The gradient of ΔB₀ along z was calculated for each time point, resulting in 60 $G_z$ maps. For each voxel, a linear regression was performed between the latter and the associated pressure measurements (*P(t)*):

$$G_z(t) = G_{z,static} + RIGO_{z,max} \cdot P(t) \tag{3}$$

The resulting *G*$_{z,static}$ and *RIGO*$_{z,max}$ images were resampled and registered to the axial MGRE scan. The *Shimming Toolbox* was used to define an ROI (called the "shimming ROI") centered along the midline of the phantom, and the average *G*$_{z,static}$ and *RIGO*$_{z,max}$ values within the ROI were measured (<*G*$_{z,static}$>$_{ROI}$ and <*RIGO*$_{z,max}$>$_{ROI}$) for each slice and written to a text file (the "shim coefficients" file) that was transferred back to the scanner.

Data Acquisition: Shimming Experiment

Once the shim coefficients (<*G*$_{z,static}$>$_{ROI}$ and <*RIGO*$_{z,max}$>$_{ROI}$) for each slice were transferred to the scanner, two additional axial MGRE ("shimming") scans were acquired. For these two scans, the "shimming" functionality of the custom sequence was activated. This will ensure that the sequence automatically retrieves the "shim coefficients" file, continuously monitors the pressure measured by the respiratory bellows, and executes the shimming scenario that the user selects.

- The first shimming condition was "static z-shimming". Here, the sequence used the <*G*$_{z,static}$>$_{ROI}$ values for each slice to compute the appropriate *G*$_{z,corr}$ values so that through-plane field inhomogeneities could be corrected for on a slice-by-slice basis. The moment of *G*$_{z,corr}$ applied prior to each echo should be equal to -<*G*$_{z,static}$>$_{ROI}$·Δ*TE*.

- The second shimming condition was "realtime z-shimming". The <*G*$_{z,static}$>$_{ROI}$ and <*RIGO*$_{z,max}$>$_{ROI}$ values for each slice were read by the sequence at the start of the acquisition. At the start of each RF excitation the sequence will retrieve the bellows' pressure (*P(t')*), so that the average gradient of the magnetic field along the slice-select axis ($<G_z>$), for the slice in question can be predicted: $<G_z> = <G_{z,static}>_{ROI} + <RIGO_{z,max}>_{ROI} \cdot P(t')$. Each $G_{z,corr}$ will have a moment equal to $-<G_z> \cdot \Delta TE$ and correct for both static and respiration-induced through-plane field inhomogeneities.

*In-Vivo*

Data Acquisition: Training Session

In-vivo scans were acquired in 12 healthy human participants on a 3 T Siemens Prisma-fit scanner using a 64-channel head-neck coil. Informed consent was given prior to the scanning session (study approved by the Comité d'éthique de la recherche du Regroupement Neuroimagerie Québec). Respiratory bellows were strapped to the volunteers' chests and connected to Siemens' physiological monitoring unit (PMU) system.

A 2D T1w anatomical axial scan was first acquired (TR = 15 ms, TE = 3ms, matrix = 256 x 256, image resolution = 0.9 x 0.9 mm$^2$, 22 slices with 3 mm slice thickness, FA = 20$^0$, in-plane acceleration factor 2).

Following the T1w scan, the scanner's shim volume was manually set to cover the area of interest in the spinal cord. Three iterations of volume-specific shimming were performed to ensure that a stable shim was applied. Next, a dual-echo field mapping sequence (TR = 35 ms, TE$_1$ = 2.14 ms, TE$_2$ = 4.6 ms, matrix = 128 x 88, ⅞ phase partial Fourier factor, image resolution = 2.2 x 2.2 mm$^2$, 5 slices with 2.2 mm slice thickness, FA = 20$^0$) was used to acquire 30 sequential, sagittal, volumes oriented along the midline of the spinal cord. The field mapping sequence was planned on the high-resolution T1w scan to ensure that the sagittal slices of the field mapping sequence covered the SC throughout C6/C7. The PMU signals read into the scanner were automatically written to a log file at the start of the field mapping series, along with the log file start and stop times.

The table position, the scanner's shim adjustment volume, and the shim values used for the dual-echo field mapping sequence were kept constant throughout all the scans that were subsequently acquired. An axial ("no shim") MGRE scan was acquired next (TR = 189 ms, 6 echoes with 2.2 ms inter-echo spacing, $TE_1$ = 2.3 ms, BW = 800 Hz/px, 12 slices contiguous slices of 3 mm thickness, matrix = 128 x 64, image resolution = 2.2 x 2.2 $mm^2$, FA = $25^0$, A/P phase encoding), using a custom sequence. The central slice was oriented perpendicularly to the spinal cord at C6/C7. The protocol for this scan was designed so as to maximize the number of slices, and consequently, the coverage of the spinal cord, while maintaining a TR time that is long enough to include multiple echoes but short enough to not lead to excessive ghosting (which would limit our ability to reliably assess signal changes in the SC).

Data Analysis: *Shimming Toolbox*

Once the data were transferred to an external computer, the *Shimming Toolbox* was used for analysis. The same analysis procedure (summarized in Figure 3) outlined for the phantom experiment was followed, except that for in-vivo data, the gradient of $\Delta B_0$ was calculated along all three axes (x, y and z) for each time point, resulting in 30 $G_x$, $G_y$, and $G_z$ gradient maps. For each voxel, a linear regression was performed between the latter and the associated pressure measurements (*P(t)*):

$$G_i(t) = G_{i,static} + RIGO_{i,max} \cdot P(t) \qquad [4]$$

The resulting $G_{i,static}$ and $RIGO_{i,max}$ images were resampled and registered to the axial MGRE scan. The spinal cord toolbox (SCT) (De Leener et al. 2017), which is integrated within the *Shimming Toolbox*, was used to create a cylindrical ROI of 21 mm diameter which was centered on the SC in the MGRE scan (see Figure 3B). The motive for selecting this ROI shape and size is that we wanted to get a representative measure of the average field gradients within the approximate area corresponding to the SC. Lastly, the average $G_{i,static}$ and $RIGO_{i,max}$ values within the ROI were measured (<$G_{i,static}$>$_{ROI}$ and <$RIGO_{i,max}$>$_{ROI}$) for each slice and written to a text file (the "shim coefficients" file) that was transferred back to the scanner.

Data Acquisition: Shimming Experiment

Once the shim coefficients ($<G_{i,static}>_{ROI}$ and $<RIGO_{i,max}>_{ROI}$) for each slice were transferred to the scanner, four additional axial MGRE shimming scans were acquired:

- static z-shimming (see phantom shimming acquisition)

- static xyz-shimming: Here, the sequence used $<G_{x,static}>_{ROI}$, $<G_{y,static}>_{ROI}$, and $<G_{z,static}>_{ROI}$ values to compute $G_{x,corr}$, $G_{y,corr}$, and $G_{z,corr}$ values so that both through-plane and in-plane field inhomogeneities could be corrected for.

- realtime z-shimming (see phantom shimming acquisition)

- realtime xyz-shimming: Before exciting each slice, the sequence will retrieve the bellows' pressure ($P(t')$), and the average gradient of the magnetic field along the slice-select and in-plane axes ($<G_x>$, $<G_y>$, and $<G_z>$), for that slice, is predicted. $G_{x,corr}$, $G_{y,corr}$, and $G_{z,corr}$ will have moments equal to $-<G_x>·\Delta TE$, $-<G_y>·\Delta TE$, and $-<G_z>·\Delta TE$ and correct for both static and respiration-induced through-plane and in-plane field inhomogeneities.

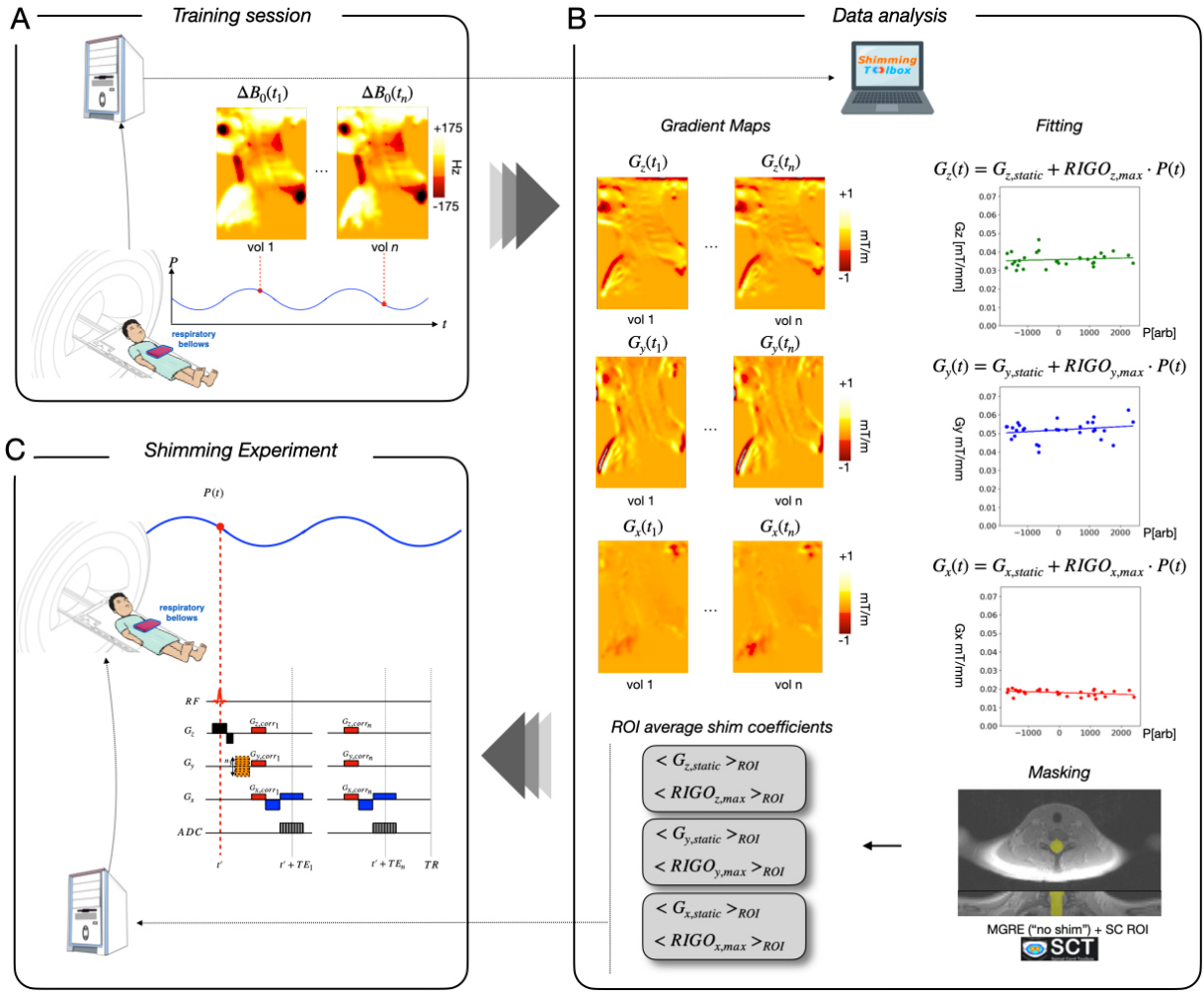

**Figure 3:** (A) Training session: a time series of ΔB$_0$ maps are acquired while simultaneously measuring and recording pressure readings obtained from bellows, (B) Data analysis: $G_i$ (i=x,y,z) map time series are computed and voxel wise linear regression of $G_i$ vs. pressure readings are performed. $G_{i,static}$ and $RIGO_{i,max}$ are averaged in an SC ROI and transferred to the scanner, (C) shimming acquisition with a modified MGRE sequence and realtime monitoring of pressure readings.

Post-processing

For each volunteer, the MGRE acquisitions were registered to the PAM50 SC template (De Leener et al. 2018) as follows: (1) The T1w scan was registered to the PAM50 template using SCT (De Leener et al. 2017) for each volunteer. (2) Each one of the MGRE acquisitions was averaged across echo times to provide a robust registration target and each averaged MGRE acquisition (i.e.: time averaged no shim, time averaged static z-shim, etc.) was registered onto the T1w scan. (3) The T1w-to-PAM50 transformation was concatenated with the MGRE-to-T1w transformation and the resulting transformation (MGRE-to-T1w-to-PAM50) was used to register the MGRE acquisitions for each shim condition (i.e.: no shim, static z-shim, etc.) to the PAM50 template.

The mean image across volunteers was computed for each echo and shim condition. Then, the SC segmentation of the PAM50 template was used to extract the average signal for each slice and echo time for every shim condition from the volunteer-averaged, registered MGRE scans. One volunteer (acdc_165) was excluded from this mean image due to poor coregistration results.

**Results**

*Phantom*

The results of our phantom validation experiment are summarized in Figure 4 and Table 1. In Figure 4 we show MGRE images with no dynamic shimming ("no shim"), static z-shimming ("static z-shim") and realtime z-shimming ("rt z-shim") for TE1 and TE2. Images show a recovery of signal loss going from "no shim" to "static z-shim" and from "static z-shim" to "rt z-shim". This is further supported by the data in Table 1, which includes the median within-ROI signal across slices for each echo time and each shimming condition.

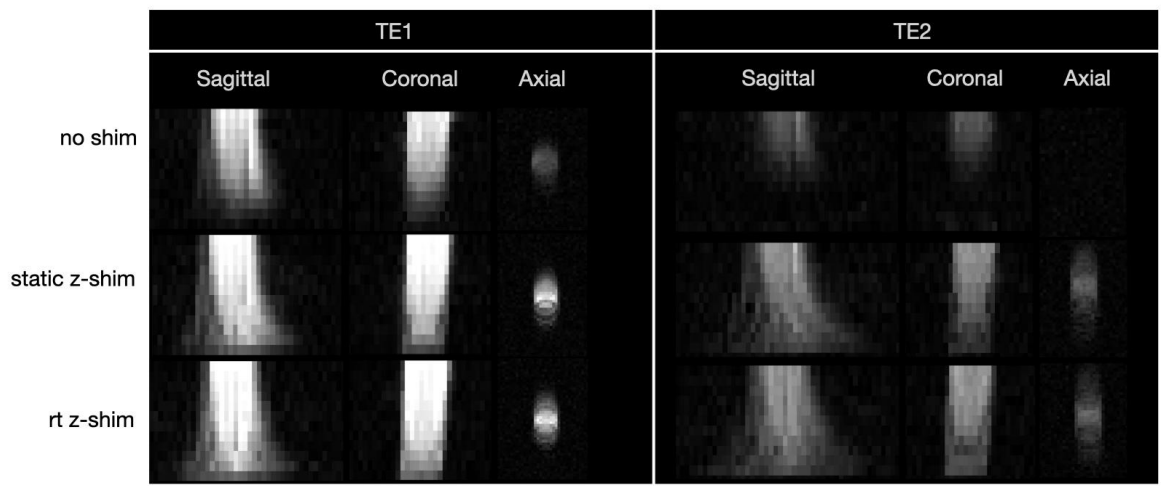

**Figure 4: MGRE images obtained in a small test tube at TE1 and TE2 using no dynamic shimming ("no shim"), static z-shimming ("static z-shim"), and realtime z-shimming ("rt z-shim").**

|  | TE1 | TE2 | TE3 | TE4 |
|---|---|---|---|---|
| No shim | 322 | 51 | 16 | 13 |
| Static z-shim | 458 | 208 | 56 | 40 |
| Realtime z-shim | 471 | 233 | 76 | 55 |

**Table 1: Median across-slice ROI-averaged MGRE signals (a. u.) for each echo time and shim condition.**

*In-Vivo*

Figure 5 includes representative MGRE images (at TE6) for a single volunteer for each shim condition, in the volunteer's native space. One can visually appreciate a gradual improvement in image quality going from the "no shim" condition to "rt z-shim" and/or "rt xyz-shim", with the exception of static xyz-shimming. Signal is recovered within the spinal cord (as indicated by the red arrows in the axial images) and the signal along the spinal cord becomes more uniform (red arrows in the coronal images).

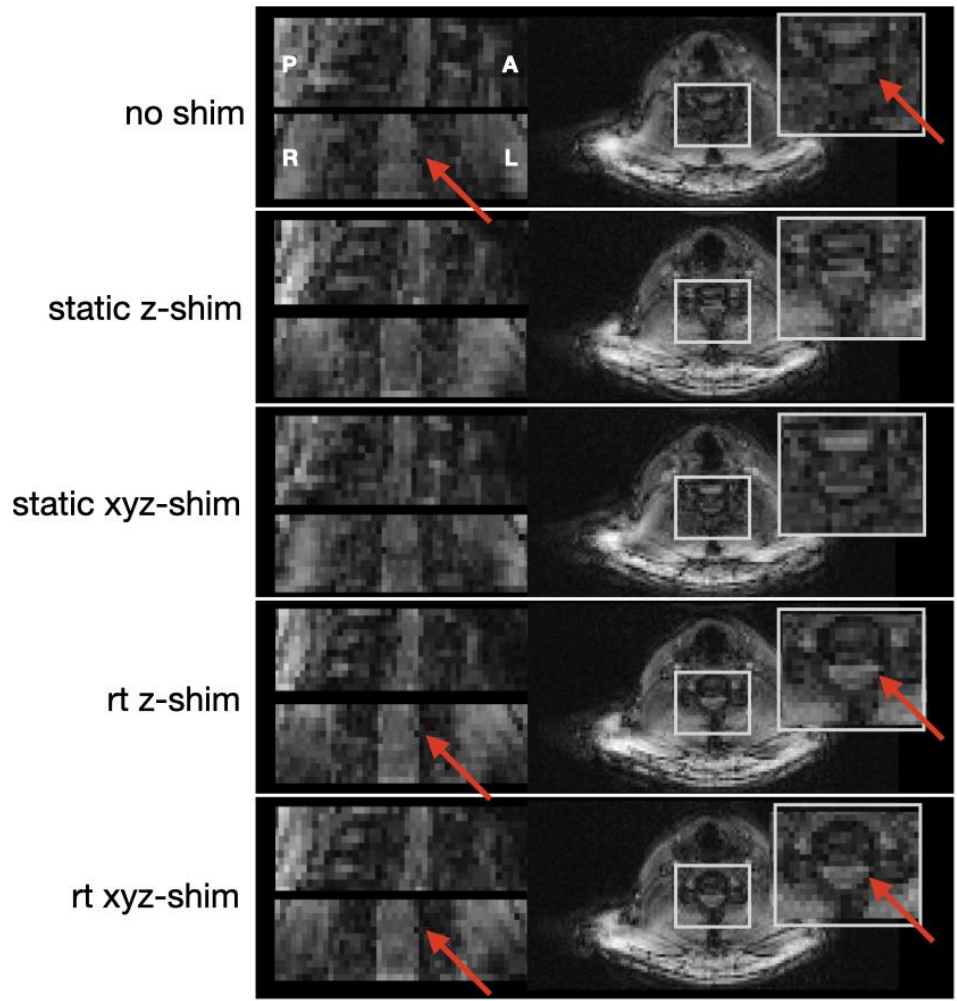

**Figure 5: Sagittal, coronal and axial views of MGRE scans (TE6) corresponding to a single volunteer with no slice wise shimming ("no shim"), static z-shimming ("static z-shim"), static xyz-shimming ("static xyz-shim"), realtime z-shimming ("rt z-shim"), and realtime xyz-shimming ("rt xyz-shim"). The red arrows indicate regions in which the improved signal homogeneity is apparent in MGRE images acquired with dynamic shimming, compared to no slice-wise shimming.**

Figure 6 contains violin plots (a box plot overlaid on a kernel density estimate (KDE)) of the mean ROI signal intensity in each slice, averaged across all volunteers. These are shown for each echo time, and each shim condition. All of the data were normalized to the "no shim" group-averaged signal at TE1, averaged across slices. Compared to the "no shim" condition, the distributions for the "static z-shim" and "rt xyz-shim" appear to be more narrowly distributed, particularly at later echo times, suggesting that the signal across slices becomes more uniform/less variable when static z-shimming or realtime xyz-shimming are used. The mean and standard deviation (std) obtained from these distributions are summarized in Table 2. A Shapiro Wilks test revealed that all of the distributions depart significantly from normality ($p<0.05$). Based on this finding, we proceeded to test for significant differences between the "no shim" condition and subsequent "shimmed" conditions at each echo time using a two-sample Kolmogorov-Smirnov test (results summarized in Figure 6). Notably, the static xyz-shimmed and realtime z-shimmed distributions at TE6 did not appear to show a significant difference relative to the "no shim" distribution. Individual p-values for all statistical tests were adjusted for multiple comparisons using the Bonferroni-Holm method. Given the exploratory nature of this work, an *a priori* power analysis was not conducted. However, an *a posteriori* sensitivity power analysis indicates that at least 90 subjects would be required to detect a standardized mean difference (Cohen's *d*) of 0.3 between two shimming conditions ($\alpha$ = 0.05, power = 0.80).

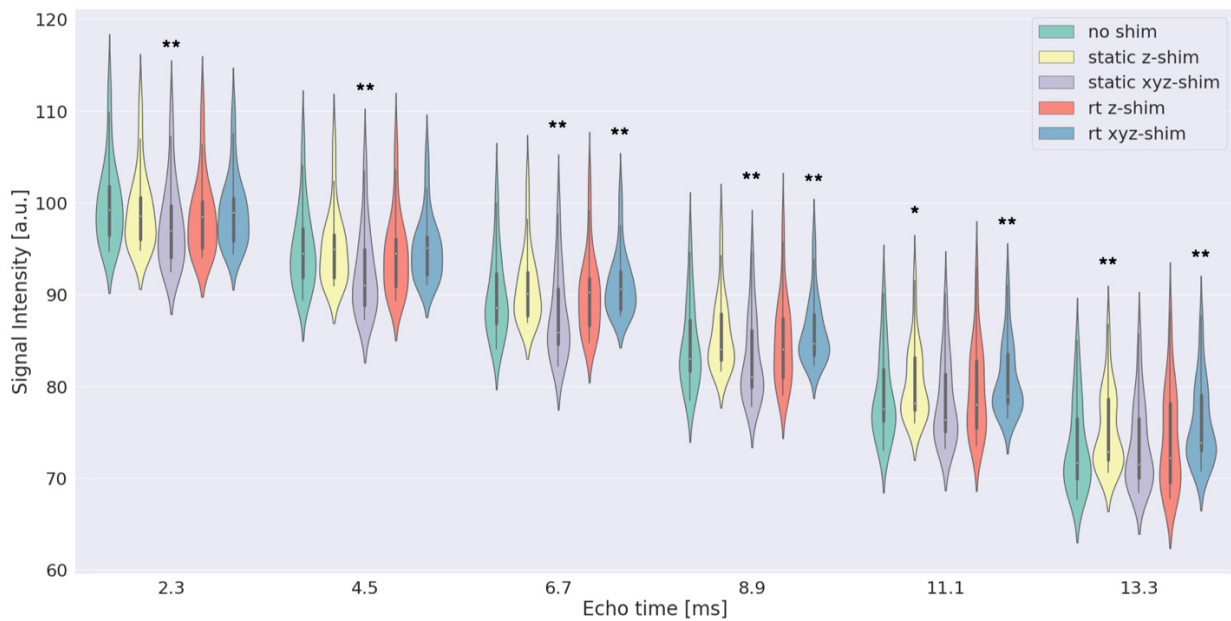

**Figure 6: Violin plot showing the distributions for "no shim" MGRE data (green), static z-shimmed MGRE data (yellow), static xyz-shimmed MGRE data (purple), realtime z-shimmed MGRE data (red), and realtime-xyz shimmed MGRE data (blue) at each echo time. The violin plots show a box plot overlaid on a kernel density estimate of within-slice SC ROI-averaged signal intensities across slices and volunteers. The white dot within the box plot indicates the median of the distribution, while the black rectangle encompasses the 25th to the 75th percentiles, and the endpoints of the lines coincide with the upper**

and lower adjacent values. * *p* < 0.1, ** *p* < 0.05. The individual p-values were adjusted for multiple comparisons and can be found in our interactive notebook (see Data/Code Sharing section).

In Table 2 we list the mean ± std for each echo time and shim condition. We then computed the percent change in the mean and the std for each shim condition (at each echo time), relative to the "no shim" condition, and took the mean of those values, across echo times. Those two results are listed in the last column of Table 2 (percent change in the mean / percent change in the std). Static z-shimming, realtime z-shimming and realtime xyz-shimming led to increases in the mean spinal cord signal (when looking at the average across echo times), compared to the "no-shim" condition. Both static z-shimming and realtime xyz-shimming also led to decreases in the spread of signals along the spinal cord (std) (when looking at the average across echo times).

|  | TE1 | TE2 | TE3 | TE4 | TE5 | TE6 | Mean % change (in the mean/ in the std) |
|---|---|---|---|---|---|---|---|
| No shim | 100.0±4.6 | 95.0±4.4 | 89.7±4.4 | 84.3±4.5 | 78.8±4.6 | 73.2±4.6 |  |
| Static z-shim | 99.1±4.2 | 95.1±4.0 | 90.8±4.0 | 85.7±3.9 | 80.3±4.0 | 75.1±4.2 | 1.1 / -10.5 |
| Static xyz-shim | 98.5±4.6 | 94.4±4.6 | 90.0±4.7 | 84.9±4.4 | 79.5±4.5 | 74.2±4.5 | -1.6 / 1.1 |
| Realtime z-shim | 97.6±4.3 | 92.4±4.3 | 87.7±4.3 | 82.9±4.6 | 78.2±4.9 | 73.4±5.3 | 0.2 / 2.4 |
| Realtime xyz-shim | 99.1±3.9 | 95.1±3.6 | 91.2±3.4 | 86.0±3.5 | 80.7±3.8 | 75.8±4.3 | 1.5 / -17.1 |

**Table 2: Mean ± std for each echo time and shim condition. The column on the right represents the mean across echo times of the percent change in the mean and std between each shim condition and the "no shim" condition.**

In Figure 7 we show group-averaged MGRE TE6 images for each shim condition. Here we observe a subtle, but gradual improvement in image quality going from the "no shim" condition (where no dynamic shimming was used) to "rt xyz-shim" (where realtime dynamic shimming was used along all three axes). Signal is recovered within the spinal cord (as indicated by the red arrows in the axial images) and the signal along the spinal cord becomes more uniform (red arrows in the coronal and sagittal images). The axial images correspond to a slice that is superior to the C6/C7 intervertebral disk, coinciding with an area in which strong magnetic field gradients are expected.

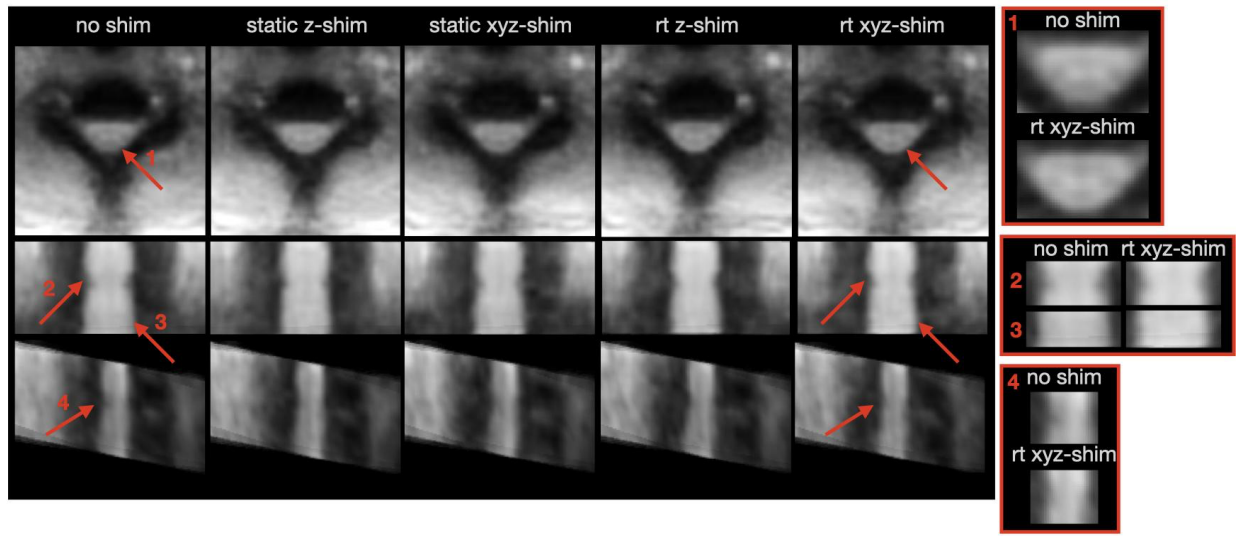

**Figure 7: Group averaged MGRE TE6 images for all 5 shimming conditions. Top: axial, Middle: coronal, Bottom: sagittal. The axial images correspond to a slice that is superior to the C6/C7 intervertebral disk. The red arrows indicate regions in which the improved signal homogeneity is apparent in MGRE images acquired with dynamic shimming, compared to no slice-wise shimming. These areas have been zoomed in on the right to facilitate visual comparisons.**

The results shown in Figure 7 are quantified in Figure 8, where the group-averaged spinal cord signal is shown along corresponding PAM50-template (De Leener et al. 2018) spinal cord levels in MGRE images for TE6. Here we observe that realtime xyz-shimming ("rt xyz-shim") signals are highest along most of the spinal cord and have the least variability. The "no shim" signals lead to the greatest variability across slices and the greatest signal loss for slices that are closest to the lungs (where more signal loss is expected).

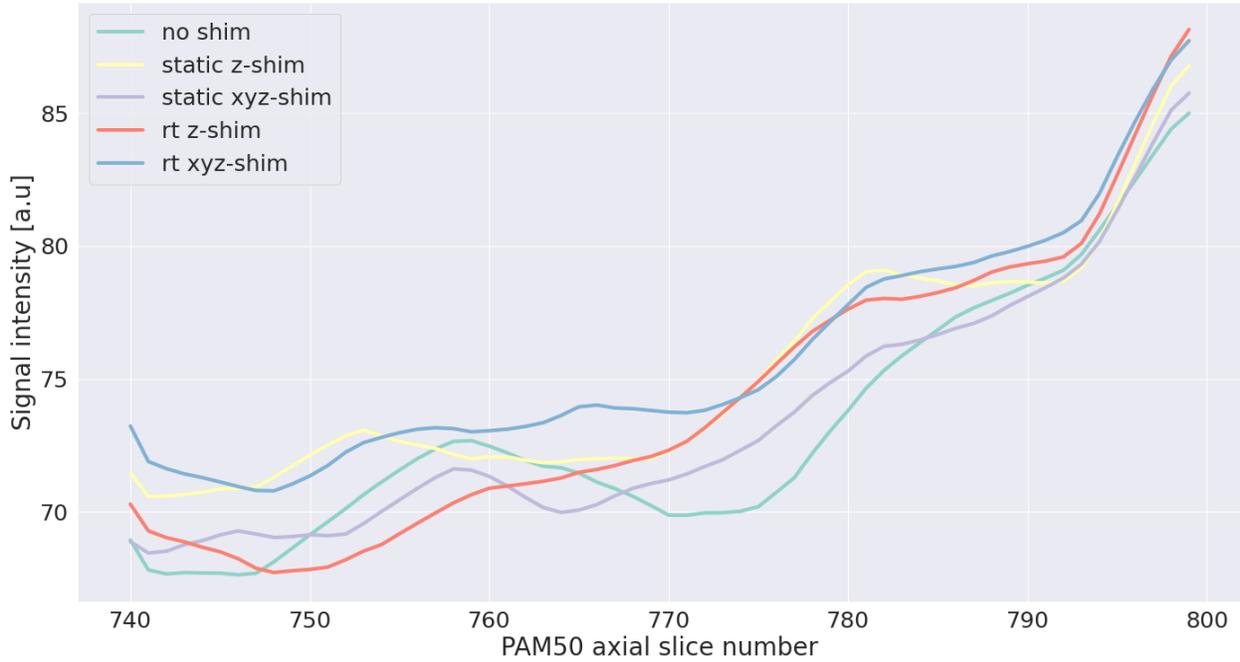

Figure 8: Group averaged mean spinal cord signals in MGRE TE6 (13.3 ms) images for all 5 shimming conditions (green: "no shim", yellow: "static z-shim", purple: "static xyz-shim", red: "rt z-shim", blue: "rt xyz-shim"), across slices. The indices on the x-axis refer to the PAM50 template (De Leener et al. 2018) slice index. Slice numbers increase along the foot-to-head direction and cover levels C7 to C6.

In order to attempt to disentangle the apparent improvement in image quality with realtime xyz-shimming, but not with static xyz-shimming, we analyzed the variability of $G_{i,static}$ and $RIGO_{i,max}$ values within the SC ROI that was used to calculate $<G_{i,static}>_{ROI}$ and $<RIGO_{i,max}>_{ROI}$. We computed the std of $G_{i,static}$ and $RIGO_{i,max}$ values within the SC ROI for each slice, and show the distribution of those values (in the form of a KDE) across slices and subjects in Figure 9. In Figure 9A we see that the $G_{y,static}$ has a greater spread in values within the SC ROI, compared to $G_{z,static}$ and $G_{x,static}$. This trend is still visible in Figure 9B but is less pronounced. The processed gradient field maps were not properly transferred for one subject (acdc_156) and are therefore excluded from Figure 9.

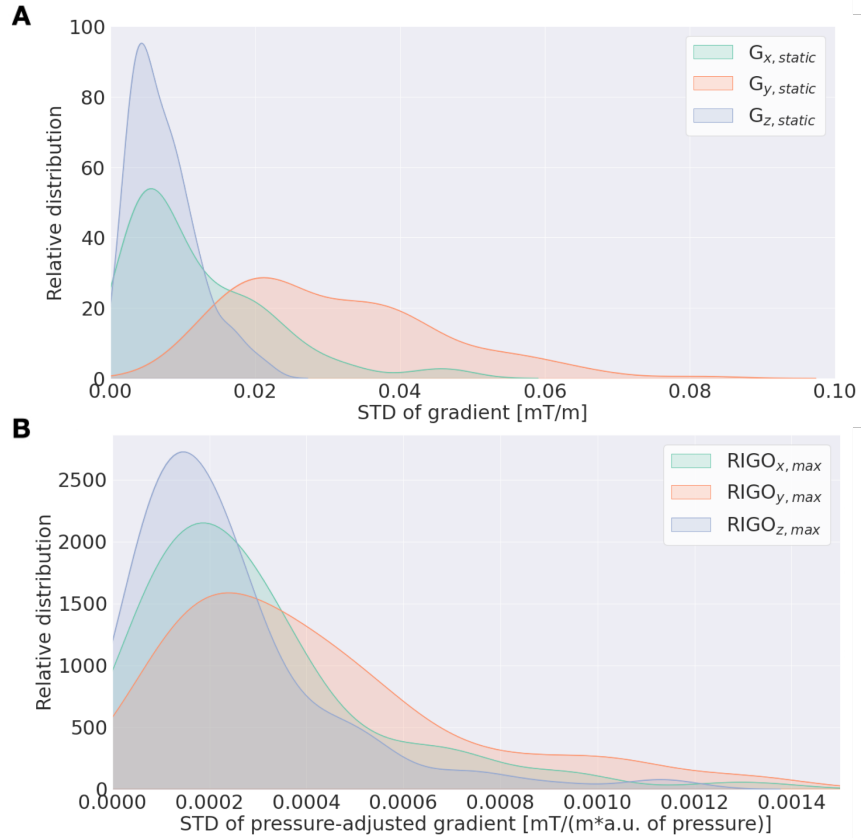

**Figure 9:** KDE of the standard deviation (STD) of A: $G_{i,static}$ within the SC ROI across slices and subjects, B: $RIGO_{i,max}$ within the SC ROI across slices and subjects.

**Discussion and Conclusions**

The SC is among the most difficult-to-image structures in the body using MRI (Martin et al. 2016), in part due to the magnetic field inhomogeneities at the interface between bone, intervertebral disks, and cerebrospinal fluid. Of the many requirements for high-quality SC imaging is advanced shimming. One approach to shimming in the SC region is to apply linear compensation gradients along the slice-select axis to compensate for through-plane field variations, and to change these compensation gradients on a per-slice basis (z-shimming (Finsterbusch, Eippert, and Büchel 2012)). This method has been found to significantly increase signal intensities in the SC and reduce signal intensity variations in $T_2^*$-weighted EPI axial acquisitions of the cervical spinal cord at 3 T. In this work, we carried out a conceptual replication (Nosek and Errington 2020) study wherein we implemented z-shimming within an MGRE sequence. We observed a small increase (1.1%) in SC signal intensity (as measured by the mean across slices) and a more prominent decrease (10.5%) in signal variability across slices (in the form of a decrease in the std of the SC signal across slices). Some important differences between the two studies are the acquisition parameters and the approach to determining the appropriate compensation gradients. With regard to the former; the echo-time used in the EPI acquisitions in (Finsterbusch, Eippert, and Büchel 2012) was 44 ms (the maximum echo time used in this study was 13.5 ms) and the coverage of the slices ranged from C5 to T1 (whereas the C6/C7 levels were covered by our scans). The longer echo time is expected to lead to greater spin dephasing and consequent signal loss in the case of the EPI scans, although this should be somewhat offset by the smaller voxel volume used by Finsterbusch et al., compared to what was used for our scans (1 x 1 x 5 mm$^3$ vs. 2.2 isotropic). The motivation behind our protocol choice was that we sought to acquire closely spaced echoes in order to reflect the acquisition parameters that may be of interest in a quantitative $T_2^*$ study. Measurements at echo times beyond 13.5 ms were not possible due to the presence of severe ghosting when TR times in excess of ~ 200 ms were used, which limited our ability to reliably measure signal changes in the spinal cord with and without shimming. The latter was mitigated in (Finsterbusch, Eippert, and Büchel 2012) using spatial saturation bands. Unfortunately, these were not available in the FLASH-based MGRE sequence that we used for our study. In (Finsterbusch, Eippert, and Büchel 2012), the authors observed the greatest signal changes along the inferior measured levels, whereas superior measured levels (which coincide with those measured in this study) had less of a signal gain when z-shimming was used (observations based on visual inspection of figures in (Finsterbusch, Eippert, and Büchel 2012)). In terms of the approach to determining the appropriate compensation gradients, (Finsterbusch, Eippert, and Büchel 2012) determined these by acquiring axial EPI scans with a range of differing compensation gradients. The compensation gradient that led to a maximum signal for each slice was then used in a subsequent (z-shimmed) EPI scan. In our approach, we determined optimal compensation gradients from field maps that were acquired prior to our z-shimmed MGRE scan. Inaccuracies in our dual-echo-measured field maps are a potential source of error when using our approach. The 2.46 ms echo spacing that was used for our $\Delta B_0$ mapping sequence corresponds to a bandwidth of ~ 200 Hz. Recent work by Islam et al. (Islam et al. 2019) reports static frequency offsets ranging from -100 to -250 Hz at 3 T in the cervical spinal cord and in (Verma and Cohen-Adad 2014), respiration-induced frequency offsets were found to reach 74 Hz at 3 T at the C7 level. Given that a direct correspondence between measured frequency offset and cervical cord level was not given in (Islam et al. 2019), it is difficult to estimate whether the bandwidth of our Δ$B_0$ mapping sequence is sufficient for the purposes of our study. There is also the possibility that the assumption of linearity in through-slice field variations does not hold in the vicinity of intervertebral disks. If this is the case, the field inhomogeneities cannot be compensated perfectly, and the gradient that best approaches the field inhomogeneity may not be the gradient that provides the maximum signal intensity (due to the non-linear relationship between field inhomogeneities and the signal). In this case, the EPI measurement-based approach has the advantage that the measure to be optimized, the signal amplitude, is directly observed. Furthermore, any experimental imperfections that may cause differences between the EPI and the field map measurement, e.g., geometric displacements or slight field-inhomogeneity-induced modulations of the slice thickness, position and orientation, are inherently considered (private communication with Finsterbusch and colleagues). Interestingly, in a more recent study, by Kaptan et al. (Kaptan et al. 2022), the authors compared the EPI measurement-based approach to a field map-based approach to estimating the required z-shimming compensation gradients, such as the one which was used in our work. The authors found that the field map-based approach underperformed, compared to the EPI measurement-based approach. Notwithstanding all of these factors, we did observe a small increase in the average SC signal across slices in the C6/C7 levels and a decrease in signal variability, compared to images in which no z-shimming was used. These findings suggest that z-shimming may be beneficial in the context of MGRE sequences.

We sought to further improve-upon SC signal stability by including in-plane compensation gradients with our MGRE sequence (along the frequency encode "x" and the phase-encode "y" directions). However, these did not lead to either signal recovery or a reduction in signal variability, but rather the contrary. We hypothesized that this may be due to a lack of homogeneity of $G_{x,static}$ and $G_{y,static}$ within the SC ROI that was used to calculate their average values. For dynamic shimming to be effective, the gradient we are attempting to correct for should be as uniform as possible within the area of interest (given that a fixed compensation gradient will be used for the entire slice). In order to verify this, we measured the variability (standard deviation) of $G_{x,static}$, $G_{y,static}$ and $G_{z,static}$ within the SC ROI and found that $G_{y,static}$ values had a greater within-ROI spread, compared to that of $G_{x,static}$ and $G_{z,static}$. We believe that this may explain the poorer performance of static xyz-shimming.

In addition to "static z-shimming" and "static xyz-shimming", we attempted to correct for respiration-induced magnetic field gradients by adjusting the compensation gradients throughout the image acquisition process in order to reflect the respiratory state. The feasibility of our proposed realtime compensation scheme was validated using a phantom setup which successfully demonstrated the ability of realtime z-shimming to recover signal loss due to time-varying $\Delta B_0$ offsets. While our phantom experiment demonstrated signal recovery in the context of time-varying $\Delta B_0$ with static z-shimming, further recovery was made possible with realtime z-shimming. In-vivo, realtime compensation was effectuated for both z-shimming and xyz-shimming. While both of these scenarios led to increases in the mean of the signals across the measured slices, only realtime xyz-shimming led to a reduction in the variability of signals across the measured slices (17.1%, the greatest among all of the tested shim conditions). We did not observe as big of a discrepancy in the spread of $RIGO_{x,max}$, $RIGO_{y,max}$ and $RIGO_{z,max}$ values within the SC ROI, compared to their static counterparts. It may be that the realtime compensation offered by realtime xyz-shimming more than overcomes the potential spin dephasing experienced by spins within the SC ROI due to mismatched $G_{y,static}$ and $G_{y,corr}$ values that could arise due to a lack of uniformity of $G_{y,static}$ values within the SC ROI. It should be noted that throughout all of the realtime experiments, the scanner operator kept an eye on the respiratory trace (which was displayed on the scanner console) in order to ensure that the subject maintained a regular breathing pattern throughout the imaging sessions. However, in the case of acquisitions that are shorter than the one respiratory period, it is more difficult to ensure that the subject is breathing normally throughout the acquisition. In this case, it is possible that the assumptions made with regard to the relationship between $\Delta B_0$ and respiration do not always hold. This would be expected to impact the efficacy of realtime dynamic shimming. A limitation of our experimental design is the time taken to acquire a single $\Delta B_0$ map. We sought to both minimize the $\Delta B_0$ map acquisition time (TR = 35 ms, 77 phase encoding steps) and maximize the number of slices (3 sagittal slices), so that $B_0$ field gradients could be accurately measured. Gradient estimates along all three cartesian axes are required for gradient correction along any one axis of an arbitrarily-orientated MGRE scan. The estimation of gradients along the patient left-right direction requires a minimum of 2 slices in the $\Delta B_0$ map. Here we acquired 3 so that we could use the central difference approach and improve gradient estimation accuracy. However, the resulting ~ 2.7 s acquisition time is long considering the ~ 5-6 s respiratory period observed in our volunteers during near-sleep resting states, and likely leads to some blurring effects. In the future, we will explore the possibility of accelerating our $\Delta B_0$ map acquisition using parallel imaging (which is not currently available with the Siemens product $\Delta B_0$ mapping sequence).

To summarize, in this work we sought to verify the translatability of (static) z-shimming to a MGRE sequence and to further improve upon the method using a combination of shimming to compensate for in-plane (xyz-shimming) and time-varying (realtime) gradients. A proof-of-concept phantom experiment demonstrated the applicability of static z-shimming to a MGRE sequence and the feasibility of our proposed realtime correction for time-varying $\Delta B_0$. In-vivo scanning results showed that static z-shimming may be beneficial when scanning the cervical spinal cord region with a MGRE sequence. The longest echo time used in our acquisition was 13.6 ms. At such an echo time, we expect a conservative effect of size for dynamic shimming. With this in mind, our study is likely underpowered. Our sample size of 12 individuals coincides with the norm within our discipline (a recent study found that 96% of highly cited experimental fMRI studies had a median sample size of 12 participants (Szucs and Ioannidis, 2019)). Notwithstanding the latter, our in-vivo realtime shimming experiments indicated that further improvements in image quality can be made when using realtime xyz-shimming. Future work will focus on exploring the utility of dynamic shimming at higher field strengths, at accelerating our MGRE sequence (by using parallel acquisition methods and/or saturation bands) such that longer echo times and a greater number of slices can be acquired in a comparable TR time, and at accelerating the Δ$B_0$ mapping sequence to improve the temporal characterization of the Δ$B_0$ field.

**Data / Code sharing**

The data collected for this study are available on OSF (https://osf.io/v4tdk/). The code used for post-processing is available through an interactive notebook (https://github.com/neuropoly/realtime-dynamic-shimming/releases/tag/v1.0.2) so that most of the results from the paper can be reproduced on the cloud. Readers can also reproduce our dynamic shimming workflow using the *Shimming Toolbox* that is publicly available: https://shimming-toolbox.org/.

**Acknowledgments**

We would like to thank Cyril Tous for his assistance in developing the realtime z-shimming MGRE sequence, Jürgen Finsterbusch for discussions on z-shimming in the cervical spinal cord, Julia Palaretti for her assistance with data analysis, Mathieu Guay-Paquet for discussions on statistical analyses, and the Functional Neuroimaging Unit for the scanning facilities. This research was undertaken thanks, in part, to funding from the Canada First Research Excellence Fund through a TransMedTech Institute postdoctoral fellowship and grant (Eva Alonso-Ortiz), the Systems, Technologies and Applications for Radiofrequency and Communications (STARaCOM) through a postdoctoral scholarship (Eva Alonso-Ortiz), the Natural Sciences and Engineering Research Council through a postdoctoral fellowship (Eva Alonso-Ortiz), a MITACS Accelerate Fellowship (Daniel Papp), the Courtois Neuromod foundation, the Canada Research Chair in Quantitative Magnetic Resonance Imaging [950-230815], the Canadian Institute of Health Research [CIHR FDN-143263], the Canada Foundation for Innovation [32454, 34824], the Fonds de Recherche du Québec - Santé [28826], the Natural Sciences and Engineering Research Council of Canada [RGPIN-2019-07244], the Canada First Research Excellence Fund (TransMedTech and IVADO) and the Quebec BioImaging Network [5886, 35450].

# References

Alonso-Ortiz, E., Levesque, I. R., and Pike, G. B. 2018. "Multi-Gradient-Echo Myelin Water Fraction Imaging: Comparison to the Multi-Echo-Spin-Echo Technique." *Magn. Reson. Med.* 79 (3): 1439–46.

Barry, R. L., and Menon, R. S. 2005. "Modeling and Suppression of Respiration-Related Physiological Noise in Echo-Planar Functional Magnetic Resonance Imaging Using Global and One-Dimensional Navigator Echo Correction." *Magn. Reson. Med.* 54 (2): 411–18.

Blamire, A. M., Rothman, D. L., and Nixon, T. 1996. "Dynamic Shim Updating: A New Approach towards Optimized Whole Brain Shimming." *Magn. Reson. Med.* 36 (1): 159–65.

Constable, R. T., and Spencer, D. D. 1999. "Composite Image Formation in Z-Shimmed Functional MR Imaging." *Magn. Reson. Med.* 42 (1): 110–17.

D'Astous, A., Cereza, G., Papp, D., Gilbert, K.M., Stockmann, J., Alonso-Ortiz, E., Cohen- Adad, J., 2022. Shimming-toolbox: An open-source software toolbox for B0 and B1 shimming in MRI. Magn. Reson. Med.. Wiley. https://doi.org/10.1002/mrm.29528.

De Leener, B., Fonov, V. S., Collins, D. L., Callot, V., Stikov, N., and Cohen-Adad, J. 2018. "PAM50: Unbiased Multimodal Template of the Brainstem and Spinal Cord Aligned with the ICBM152 Space." *NeuroImage* 165: 170–79.

De Leener, B., Lévy, S., Dupont, S. M., Fonov, V. S., Stikov, N., Collins, D. L., Callot, V., and Cohen-Adad, J. 2017. "SCT: Spinal Cord Toolbox, an Open-Source Software for Processing Spinal Cord MRI Data." *NeuroImage* 145 (Pt A): 24–43.

Finsterbusch, J., Eippert, F., and Büchel, C. 2012. "Single, Slice-Specific Z-Shim Gradient Pulses Improve T2*-Weighted Imaging of the Spinal Cord." *NeuroImage* 59 (3): 2307–15.

Glover, G. H., Li, T. Q., and Ress, D. 2000. "Image-Based Method for Retrospective Correction of Physiological Motion Effects in fMRI: RETROICOR." *Magn. Reson. Med.* 44 (1): 162–67.

Islam, H., Law, C. S. W., Weber, K. A., Mackey, S. C., and Glover, G. H. 2019. "Dynamic per Slice Shimming for Simultaneous Brain and Spinal Cord fMRI." *Magn. Reson. Med.* 81 (2): 825–38.

Jenkinson, M. 2003. "Fast, Automated, N-Dimensional Phase-Unwrapping Algorithm." *Magn. Reson. Med.* 49 (1): 193–97.

Jezzard, P., and Balaban, R. S. 1995. "Correction for Geometric Distortion in Echo Planar Images from B0 Field Variations." *Magn. Reson. Med.* 34 (1): 65–73.

Juchem, C, Brown, P. B., Nixon, T. W., McIntyre, S., Rothman, D. L., and de Graaf, R. A. 2011. "Multicoil Shimming of the Mouse Brain." *Magn. Reson. Med.* 66 (3): 893–900.

Kaptan, M, Vannesjo, S. J., Mildner, T., Horn, U., Hartley-Davies, R., Oliva, V., Brooks, J. C. W., Weiskopf, N., Finsterbusch, J., and Eippert, F. 2022. "Automated Slice-Specific Z-Shimming for fMRI of the Human Spinal Cord." 2022 *Human Brain Mapping*. DOI: 10.1002/hbm.26018

Martin, A. R., Aleksanderek, I., Cohen-Adad, J., Tarmohamed, Z., Tetreault, L., Smith, N., Cadotte, D. W., et al. 2016. "Translating State-of-the-Art Spinal Cord MRI Techniques to Clinical Use: A Systematic Review of Clinical Studies Utilizing DTI, MT, MWF, MRS, and fMRI." *NeuroImage. Clin.* 10: 192–238.

Morozov, Darya, Lopez Rios, N., Duval, T., Foias, A., and Cohen-Adad, J. 2018. "Effect of Cardiac-Related Translational Motion in Diffusion MRI of the Spinal Cord." *Magn. Reson. Imag.* 50: 119-124.

Nosek, B. A., and Errington, T. M. 2020. "What Is Replication?" *PLoS Biology* 18 (3): e3000691.

Pfeuffer, J., Van de Moortele, P.-F., Ugurbil, K., Hu, X., and Glover, G. H. 2002. "Correction of Physiologically Induced Global off-Resonance Effects in Dynamic Echo-Planar and Spiral Functional Imaging." *Magn. Reson. Med.* 47 (2): 344–53.

Punchard, W. F. B., Lo, K.-M., Starewicz, P. M., and Hetherington, H. 2013. Shim insert for high-field MRI magnets. USPTO 8536870. *US Patent*, filed April 6, 2011, and issued September 17, 2013.
https://patentimages.storage.googleapis.com/11/2d/2d/71081e951ed819/US8536870.pdf

Reeder, S. B., Atalar, E., Bolster, B. D., and McVeigh, E. R. 1997. "Quantification and Reduction of Ghosting Artifacts in Interleaved Echo-Planar Imaging." *Magn. Reson. Med*. 38(3):429-439.

Ruetten, P. P. R., Gillard, J. H., and Graves, M. J. 2019. "Introduction to Quantitative Susceptibility Mapping and Susceptibility Weighted Imaging." *Br. J. Radiol.* 92 (1101): 20181016.

Stockmann, J. P., and Wald, L. L. 2018. "In Vivo B0 Field Shimming Methods for MRI at 7T." *NeuroImage* 168: 71–87.

Szucs, D., and Ioannidis, J. P. A. 2019. "Sample Size Evolution in Neuroimaging Research: An Evaluation of Highly-Cited Studies (1990-2012) and of Latest Practices (2017-2018) in High-Impact Journals." Neuroimage 221 (117164).

Topfer, R., Starewicz, P., Lo, K.-M., Metzemaekers, K., Jette, D., Hetherington, H. P., Stikov, N., and Cohen-Adad, J. 2016. "A 24-Channel Shim Array for the Human Spinal Cord: Design, Evaluation, and Application." *Magn. Reson. Med.* 76 (5): 1604–11.

Topfer, R., Foias, A., Stikov, N., and Cohen-Adad, J. 2018. "Real-Time Correction of Respiration-Induced Distortions in the Human Spinal Cord Using a 24-Channel Shim Array." *Magn. Reson. Med.* 80 (3): 935-946.

van Gelderen, P., de Zwart, J. A., Starewicz, P., Hinks, R. S., and Duyn, J. H. 2007. "Real-Time Shimming to Compensate for Respiration-Induced B0 Fluctuations." *Magn. Reson. Med.* 57 (2): 362–68.

Vannesjo, S. J., Clare, S., Kasper, L., Tracey, I., and Miller, K. L. 2019. "A Method for Correcting Breathing-Induced Field Fluctuations in T2*-Weighted Spinal Cord Imaging Using a Respiratory Trace." *Magn. Reson. Med.* 81 (6): 3745–53.

Verma, T., and Cohen-Adad, J. 2014. "Effect of Respiration on the B0 Field in the Human Spinal Cord at 3T." *Magn. Reson. Med.* 72 (6): 1629–36.

Yablonskiy, D. A., Sukstanskii, A. L., Luo, J., and Wang, X. 2013. "Voxel Spread Function Method for Correction of Magnetic Field Inhomogeneity Effects in Quantitative Gradient-Echo-Based MRI." *Magn. Reson. Med.* 70 (5): 1283–92.